\documentclass[aps,prl,reprint,superscriptaddress,nofootinbib]{revtex4-1}
\usepackage{epsfig}
\usepackage{amssymb,amsmath,amsfonts,amsthm,dsfont}
\usepackage{graphicx}
\usepackage[caption=false]{subfig}
\usepackage{float}
\usepackage{lipsum}
\usepackage{array}
\usepackage{outlines}
\usepackage[normalem]{ulem}
\usepackage{color}
\usepackage{mathtools}
\usepackage[english]{babel}
\usepackage[utf8x]{inputenc}
\usepackage[T1]{fontenc}
\usepackage{import}

\usepackage{physics}
\usepackage{qcircuit}
\usepackage{enumitem}
\usepackage{amsmath}
\usepackage{graphicx}
\usepackage[colorlinks=true, allcolors=blue]{hyperref}
\newcommand{\x}{\mathbf{x}}

\begin{document}

\title{On quantum ensembles of quantum classifiers}
\author{Amira Abbas}
\email{amyami187@gmail.com}
\affiliation{Quantum Research Group, School of Chemistry and Physics, University of KwaZulu-Natal, Durban, KwaZulu-Natal, 4001, South Africa}

\author{Maria Schuld}
\email{schuld@ukzn.ac.za}
\affiliation{Quantum Research Group, School of Chemistry and Physics, University of KwaZulu-Natal, Durban, KwaZulu-Natal, 4001, South Africa}

\author{Francesco Petruccione}
\email{petruccione@ukzn.ac.za}
\affiliation{Quantum Research Group, School of Chemistry and Physics, University of KwaZulu-Natal, Durban, KwaZulu-Natal, 4001, South Africa}
\affiliation{National Institute for Theoretical Physics (NITheP), KwaZulu-Natal, 4001, South Africa}

\begin{abstract}
Quantum machine learning seeks to exploit the underlying nature of a quantum computer to enhance machine learning techniques. A particular framework uses the quantum property of superposition to store sets of parameters, thereby creating an ensemble of quantum classifiers that may be computed in parallel. The idea stems from classical ensemble methods where one attempts to build a stronger model by averaging the results from many different models. In this work, we demonstrate that a specific implementation of the quantum ensemble of quantum classifiers, called the accuracy-weighted quantum ensemble, can be fully dequantised. On the other hand, the general quantum ensemble framework is shown to contain the well-known Deutsch-Jozsa algorithm that notably provides a quantum speedup and creates the potential for a useful quantum ensemble to harness this computational advantage.

\end{abstract}

\maketitle
\def\one{{\mathchoice {\rm 1\mskip-4mu l} {\rm 1\mskip-4mu l} {\rm
\mskip-4.5mu l} {\rm 1\mskip-5mu l}}}

\section{Introduction}
\noindent
An ensemble algorithm may be thought of as a model whose output considers the results generated from many models. The notion of ensembles dates as far back as early civilisations where majority voting was employed to make better decisions \cite{zhangma}. One of the first ensemble methods proposed was Bayesian model averaging where one integrates (i.e. averages) over all possible parameters a model may contain to derive a solution \cite{hoeting1999bayesian}. This allows for optimal predictive ability but an obvious problem is integrating over a possibly infinite parameter space, which is not always feasible \cite{madigan1994model}. This method, however, inspired the design of a quantum algorithm created to replicate an ensemble framework \cite{hybridensemble,hansen1990neural}. An exponentially large ensemble of quantum classifiers are computed in parallel, weighed according to a specified weighing scheme and their collective prediction is obtained through a single qubit measurement. The general routine is outlined in \cite{schuld2018quantum} and termed the quantum ensemble of quantum classifiers. 

In a quantum machine learning setting, there is usually interplay between classical and quantum devices such as outsourcing difficult calculations to a quantum computer \cite{kernelsmaria,kubler2019quantum,havlivcek2019supervised,benedetti2018quantum}, or classically optimising parameterised quantum algorithms \cite{4variational,romero2017quantum,farhi2014quantum,verdon2017quantum}. The quantum ensemble of quantum classifiers, however, is intended to be executed end-to-end on a quantum device and designed to harness the properties of quantum computing \cite{schuld2018supervised}. Quantum machine learning algorithms are often under scrutiny due to bold claims and high expectations, such as exponential speed ups or solving classically intractable tasks. A good first step in analysing the validity of these claims is to check if the quantum machine learning algorithm can be efficiently replicated on a classical computer. A specific implementation of a quantum ensemble, also presented in \cite{schuld2018quantum}, proposes a routine to weigh quantum classifiers by their accuracy on a training set. In this article, this ``accuracy-weighted'' scheme is shown to efficiently translate to a classical algorithm, whilst the general quantum ensemble framework is proven to contain the Deutsch-Jozsa algorithm and thus has the ability to compute a classically hard problem where quantum computing may offer an advantage \cite{DeutschJozsa,simon1997power}. 

\section{Review of the quantum ensemble algorithm}\label{review2}
\noindent Consider a dataset of the form $\mathcal{D} = \{ (\x_1, y_1), \ldots,$ $(\x_M, y_M) \}$ where $ \x_i \in R^d$ are $\ell_2$ normalised input vectors and $y_i \in \{\pm 1\}$ are their associated labels. For a new input $\tilde{\x} \in R^d$, the objective is to find its correct label $\tilde{y}$. A classifier function, $f$, may be written as follows 

\begin{equation}\label{classifier}
  y = f(\x; \theta),  
\end{equation}

\noindent with $\x \in \mathcal{X}$ and model parameters $\theta$. The parameters are then ``learned'' by fitting the model $f$ to the data, $\mathcal{D}$. The ensemble approach allows for multiple sets of parameters to be learned and this forms multiple models whose predictions are then weighed according to some optimal weighing scheme. The ensemble prediction for $\tilde{y}$ is the average
\begin{equation}\label{classicalensemble}
        \tilde{y} = \mathrm{sgn} \left(\sum_{\theta \in \mathcal{E}} w_\theta f(\tilde{\x}; \theta) \right),
\end{equation}

\noindent where $w_\theta$ weighs the models $f(\tilde{\x};\theta)$ in the ensemble $\mathcal{E}$ which are specified by $\theta$. The sign function ensures the predicted class is either +1 or -1. 

Formulating the quantum framework, suppose that parameter sets are represented by $n$-qubit states $\ket{\theta}$. A superposition of parameters $\sum_\theta \ket{\theta}$ allows for evaluation of models in parallel, whilst a state preparation routine can be used to weigh each model appropriately, thus creating the quantum ensemble. 

Following the methodology in \cite{schuld2018quantum}, a quantum routine $\mathcal{W}$ is used to change the uniform weights of model parameters in superposition to a desired weighing scheme 
\begin{equation}\label{weight}
    \mathcal{W} \ket{\x} \frac{1}{\sqrt{E}} \sum_\theta \ket{\theta} \ket{0} \to \ket{\x} \frac{1}{\sqrt{E_{\mathcal{X}}}} \sum_\theta \sqrt{w_\theta} \ket{\theta} \ket{0},
\end{equation}

\noindent where $\frac{1}{\sqrt{E_\mathcal{X}}}$ is a normalisation factor that ensures that $\sum_\theta \frac{w_\theta}{E_\mathcal{X}} = 1$ and $w_\theta$ corresponds to classical weights as depicted in Equation (\ref{classicalensemble}).

Consider a second quantum routine $\mathcal{A}$ which computes a classifier function $f(\x;\theta)$ such that
\begin{equation}\label{Aoperator}
    \mathcal{A} \ket{\x} \ket{\theta} \ket{0} \to \ket{\x} \ket{\theta} \ket{f(\x;\theta)}.
\end{equation}

\noindent The model output is stored in the last qubit $f(\x;\theta)$ where $\ket{0}$ represents class -1 and $\ket{1}$ represents class 1. $\mathcal{A}$ may also be applied to a superposition of parameter states as presented in Equation (\ref{weight}). Together, the quantum routines $\mathcal{W}$ and $\mathcal{A}$ form the quantum ensemble of quantum classifiers. The final quantum state aiming to predict the label of a new input $\tilde{\x}$ is
\[
\ket{\tilde{\x}} \frac{1}{\sqrt{E_{\mathcal{X}}}} \sum_\theta \sqrt{w_\theta} \ket{\theta} \ket{f(\tilde{\x};\theta)},
\]
and the measurement statistics of the last qubit contain the necessary ensemble prediction. Since the probability of measuring the last qubit in state $\ket{0}$ corresponds to a class prediction of -1, the class prediction probabilities are therefore,
\[
p(\tilde{y}=-1) = \sum_{\theta \in \mathcal{E}^+} \frac{w_\theta}{E_{\mathcal{X}}},
\]
and
\[
p(\tilde{y}=1) = \sum_{\theta \in \mathcal{E}^-} \frac{w_\theta}{E_{\mathcal{X}}},
\]

\noindent where $\mathcal{E}^\pm$ are subsets of $\mathcal{E}$ containing models only with predictions equal to $\pm 1$.

\section{Weighing models by their accuracy}
\noindent In \cite{schuld2018quantum}, one of the proposed weighing schemes for the quantum ensemble (referred to as q-ensemble from hereon) is to use model accuracy. In other words, more accurate models' predictions should have higher weights than those less accurate. Model accuracy $a_\theta$ is defined here as the proportion of correct classifications on the training set and the full routine to create the accuracy-weighted q-ensemble is described in \cite{schuld2018quantum}. In short, one requires an ``accuracy'' register in addition to the data, parameter and output registers
\[
\underbrace{\ket{0...0;0}}_{\mathrm{data}} \otimes \underbrace{\ket{0...0}}_{\mathrm{parameter}} \otimes \underbrace{\ket{0}}_{\mathrm{output}} \otimes \underbrace{\ket{0}}_{\mathrm{accuracy}}.
\]
First, the accuracy and parameter registers are put into a uniform superposition. Thereafter, each training pair is loaded into the data register, their model output is computed by applying the routine $\mathcal{A}$ and this is compared to the actual class label contained in the data register. The accuracy qubit is then rotated by a small amount toward $\ket{0}$ or $\ket{1}$ depending on whether the actual and predicted outputs are the same or not. After all the training pairs are loaded, the accuracy register (which is entangled with the data register) is in state
\begin{equation}\label{Atheta}
    \ket{A_\theta} = \sqrt{a_\theta}\ket{0}+\sqrt{1-a_\theta}\ket{1}.
\end{equation}
By postselecting the $\ket{0}$-branch of the accuracy qubit, then loading the new input $\tilde{\x}$ and computing the model output using $\mathcal{A}$, the final quantum state is 
\begin{equation}\label{qensemble}
    \ket{\psi} = \frac{1}{\sqrt{E_{\mathcal{X}}}}\sum_\theta \sqrt{a_\theta} \ket{\theta} \ket{f(\tilde{\x};\theta)},
\end{equation}
with measurement statistics on the output qubit corresponding directly to the ensemble classifier in Equation (\ref{classicalensemble}) with accuracy as weights. Furthermore, it was shown that the q-ensemble in Equation (\ref{qensemble}) may be constructed to contain models only with $a_\theta > 0.5$ if the assumption of a point-symmetric parameter space holds. This is an intriguing result as classical machine learning literature supports that strong ensembles are built from classifiers with $a_\theta > 0.5$ \cite{schapire2003boosting}. 

\section{Dequantising the accuracy-weighted scheme}
\noindent It turns out that the simple ``greater than 0.5 accuracy'' weighing scheme may be translated to a classical algorithm, namely, rejection sampling \cite{casella2004generalized}. 

Rejection sampling allows one to sample from a distribution $p(x)$ that is difficult to sample from, but point evaluations for a given $x$ are possible. The procedure involves choosing a distribution $q(x)$ that is easy to sample from, such as a Gaussian distribution, and scaling it by some finite factor G such that $p(x) \leq Gq(x)$  $\forall$  $x$. Next, a new sample $x'$ is drawn from $q(x)$ and a sample $u'$ is drawn from a uniform distribution in the interval $[0, 1]$. The new sample $x'$ is accepted as a sample from the original distribution if $u'Gq(x') < p(x')$ and rejected otherwise \cite{schuldbook}. The probability of acceptance may be written as \[p_{\mathrm{accept}} = \langle p(u' < \frac{p(x')}{Gq(x')}) \rangle = \langle \frac{p(x')}{Gq(x')} \rangle,\] which reduces to $\frac{1}{G}$. Returning to the accuracy-weighted q-ensemble, before postselection and ignoring the data and output registers, the quantum state is of the form
\[
\underbrace{\sum_{\theta}^{n}\frac{1}{\sqrt{E_\theta}}\ket{\theta}}_{\mathrm{parameters}} \otimes \underbrace{ \left(\sqrt{a_\theta}\ket{0}+\sqrt{1-a_\theta}\ket{1}\right)}_{\mathrm{accuracy}}.
\]

\noindent Measuring the $\ket{\theta}$ register would be equivalent to sampling from the distribution $Q = \{ \frac{1}{E_1}, \frac{1}{E_2}, ..., \frac{1}{E_n}\}$. The goal, of course, is to sample from the more complicated distribution $P = \{ \frac{a_1}{E_1}, \frac{a_2}{E_2}, ... ,\frac{a_n}{E_n}\}$. 

\noindent If we associate 
\[
Q \leftrightarrow q(x),
\] and 
\[
P \leftrightarrow p(x),
\]
then the postselection step in the accuracy-weighted q-ensemble mirrors the rejection decision in classical rejection sampling. The probability of selecting the accuracy qubit in state $\ket{0}$ for state $\ket{i}$ is $\sum_i \frac{a_i}{E_i}$. Comparing this to the rejection sampling probability of success for sample $i$ 

\[
p_{\mathrm{accept}}(i) = \langle p(u'\frac{1}{E_i} < \frac{1}{E_i} \frac{a_i}{E_i}) \rangle = \langle \frac{a_i}{E_i} \rangle, 
\]

\noindent which reduces to $\sum_i \frac{a_i}{E_i}$, shows that the two probabilities are equivalent. More concisely, if we consider a base distribution that is comprised of models' accuracies and enclose this by an easy-to-sample-from uniform distribution, by performing rejection sampling we can essentially sample from the complicated distribution of models' accuracies and reject if the sample is less than 0.5. This leaves us with a sample of models whose accuracies are greater than 0.5 from which we may obtain their weighted predictions. This directly corresponds to measurements obtained by the accuracy-weighted q-ensemble routine that postselects the $\ket{0}$-branch which only contains models with accuracy greater than 0.5 and computes their weighted prediction. 

Whilst rejection sampling is widely used in classical machine learning, its success depends largely on the task at hand and there are known issues when sampling in very high dimensions \cite{dynamicprogramming, highdimconvexity, beyer1999nearest}. As a consequence, the accuracy-weighted q-ensemble can too encounter these issues when the input data is very high-dimensional (i.e. it has many features) as demonstrated in Appendix A. That being said, this does not rule out the possibility of finding a particular high-dimensional dataset for which the accuracy-weighted scheme is successful \cite{zimek2012survey}. More importantly, there may be other weighing schemes that are more robust than the accuracy-weighted q-ensemble that cannot be translated to a classical algorithm efficiently.

\section{Linking the Deutsch-Jozsa and quantum ensemble frameworks}
\noindent The Deutsch-Jozsa algorithm provides a simple problem that a quantum computer can solve exponentially faster than a classical one \cite{DeutschJozsa,gulde2003implementation}. Consider a function $g: \{0,1\}^n \to \{0,1\}$ where $g$ is guaranteed to be either constant (all zeros or all ones) or balanced (half zeros and half ones). The goal is to determine if $g$ is constant or balanced. This would take a classical computer at worst $2^n/ 2+1$ queries of the function $g$, but using the Deutsch-Jozsa protocol on a quantum computer, one may achieve the goal with certainty in just one measurement \cite{nielsen2002quantum}. Assuming a black box quantum computer is available, the algorithm may be outlined as follows:
\begin{enumerate}[label=\roman*.]
    \item Prepare the initial state $\ket{\psi_0} = \ket{0}^{\otimes n} \ket{1}$.
    \item Apply Hadamard gates to all qubits. The quantum state becomes 
    \[
    \ket{\psi_1} = \frac{1}{\sqrt{2^n}} \sum_{x \in \{0,1\}^n} \ket{x} \left[\frac{1}{\sqrt{2}}\left[\ket{0} - \ket{1} \right] \right].
    \]
    \item \label{Uf} Next, the function $g$ is applied using a unitary transform $U_g: \ket{x,y} \to \ket{x,y \oplus g(x)}$ yielding
    \[
    \ket{\psi_2} = \sum_x \frac{(-1)^{g(x)} \ket{x}}{\sqrt{2^n}} \left[\frac{\ket{0}-\ket{1}}{\sqrt{2}} \right].
    \]
    \item Applying Hadamard gates to the first $n$ qubits then changes the quantum state to
    \[
    \ket{\psi_3} = \sum_z \sum_x \frac{(-1)^{x \cdot z + g(x)}\ket{z}}{2^n} \left[\frac{\ket{0}-\ket{1}}{\sqrt{2}} \right],
    \]
    where $x \cdot z$ is the bitwise inner product of $x$ and $z$, modulo 2.
    \item Finally, the probability of measuring the state $\ket{0}^{\otimes n}$ is
    \[
    p(0 \dots 0) = \left|\frac{1}{2^n}\sum_x (-1)^{g(x)} \right|^2,
    \]
    which is equal to 1 if $g(x)$ is constant and 0 if $g(x)$ is balanced. 
\end{enumerate} 

\noindent After having reviewed the general template for the q-ensemble, an interesting relationship between the Deutsch-Jozsa algorithm and the q-ensemble presents itself. In fact, the q-ensemble framework can be shown to contain the Deutsch-Jozsa algorithm if we choose $\mathcal{W}$ from Equation (\ref{weight}) to implement uniform weights and let $\mathcal{A} = \left(H^{\otimes n}\otimes I\right) U_g \left( I^{\otimes n}\otimes X \right) \left(I^{\otimes n}\otimes H \right)$ where $U_g$ is again a unitary implementing the function $g$. We will label these specific routines $\mathcal{W}_{DJ}$ and $\mathcal{A}_{DJ}$ and the function $g$ may now be thought of as a combination of changing the uniform weights to a desired weighing scheme $w_\theta$ and computing the model function $f$ from Equation (\ref{classifier}). We make this explicit in the following steps:

\begin{enumerate}
    \item Prepare the initial state $\ket{\psi_0} = \ket{0}^{\otimes n} \ket{1}$ such that $n$ is equal to the bit length of the parameters $\theta$. We assume that the data register is empty $(\ket{\x} = \emptyset)$.
    
    \item Apply $\mathcal{W}_{DJ}$
    \[
    \ket{\psi_1} = \frac{1}{\sqrt{2^n}} \sum_{\theta \in \{0,1\}^n} \ket{\theta} \frac{1}{\sqrt{2}}(\ket{0}-\ket{1}),
    \]
    where the computational basis states of $\ket{\theta}$ represent model parameters of bit length $n$ in a uniform superposition. 
    
    \item Apply $\mathcal{A}_{DJ}$
    \[
    \ket{\psi_2} = H^{\otimes n} \frac{1}{\sqrt{E_\mathcal{X}}} \sum_{\theta \in \{0,1\}^n} \sqrt{w_\theta} \ket{\theta} \ket{f},
    \]
    where $f$ is the model function output stored in the last qubit. $\mathcal{A}_{DJ}$ applies a Hadamard and X-gate to the last qubit, putting it into state $\ket{0}$. Then $U_g$ implements the non-uniform weights and computes the model function. Finally, Hadamard gates are applied to the first $n$ qubits - the reason for this is to fully incorporate the q-ensemble into the Deutsch-Jozsa framework. They will be ``uncomputed'' in the next step.
    
    \item Apply $n$ Hadamard gates. This yields the desired state for the q-ensemble
    \[
    \ket{\psi_3} = \frac{1}{\sqrt{E_\mathcal{X}}} \sum_{\theta \in \{0,1\}^n} \sqrt{w_\theta} \ket{\theta} \ket{f}.
    \]
    
    \item Lastly, the probability of measuring $\ket{f}$ in state $\ket{0}$ is
    \[
    p(f = 0) = \sum_{\theta \in \mathcal{E}^+} \frac{w_\theta}{E_{\mathcal{X}}},
    \]
    which corresponds to a prediction of $\tilde{y} = -1$ for the q-ensemble, where $\mathcal{E}^+$ is defined as before in the review.
\end{enumerate} 

\noindent Through congruence of steps i. to v. with steps 1. to 5. we have demonstrated that the q-ensemble contains a classically hard computation that is amenable to a quantum advantage. Finding a routine that is useful for machine learning and harnesses this quantum advantage opens doors for further research.

\section{Conclusion}
\noindent 
Ensemble methods have established themselves firmly as successful candidates for machine learning tasks in the classical realm \cite{zhang}. Quantum ensembles, on the other hand, have yet to make such an affirmation. In this article we review the notion of quantum ensembles and show that a proposed implementation may be efficiently dequantised and studied. In contrast, the general quantum ensemble framework demonstrates the ability to compute a classically intractable task by containing the Deutsch-Jozsa algorithm. This Deutsch-Jozsa reduction, however, does not necessarily represent a quantum ensemble that is useful for machine learning. As a result, the problem of finding a meaningful quantum ensemble for machine learning purposes that makes use of this quantum advantage is still open. This article offers insight on translating proposed schemes to classical sampling methods and illustrates that the potential to find a classically intractable ensemble prone to a quantum speedup exists, which brings us closer to finding meaningful machine learning applications using ensemble techniques on quantum computers. 

\section{Acknowledgements}
\noindent This work is based upon research supported by the South African Research Chair Initiative of the Department of Science and Technology and National Research Foundation.

\bibliographystyle{unsrt}
\bibliography{ensemblebib2}

\begin{thebibliography}{10}

\bibitem{zhangma}
Cha Zhang and Yunqian Ma.
\newblock {\em Ensemble machine learning: methods and applications}.
\newblock Springer, 2012.

\bibitem{hoeting1999bayesian}
Jennifer~A Hoeting, David Madigan, Adrian~E Raftery, and Chris~T Volinsky.
\newblock Bayesian model averaging: a tutorial.
\newblock {\em Statistical Science}, pages 382--401, 1999.

\bibitem{madigan1994model}
David Madigan and Adrian~E Raftery.
\newblock Model selection and accounting for model uncertainty in graphical
  models using occam's window.
\newblock {\em Journal of the American Statistical Association},
  89(428):1535--1546, 1994.

\bibitem{hybridensemble}
Arjun Chandra and Xin Yao.
\newblock Evolving hybrid ensembles of learning machines for better
  generalisation.
\newblock {\em Neurocomputing}, 69(7-9):686--700, 2006.

\bibitem{hansen1990neural}
Lars~Kai Hansen and Peter Salamon.
\newblock Neural network ensembles.
\newblock {\em IEEE Transactions on Pattern Analysis \& Machine Intelligence},
  (10):993--1001, 1990.

\bibitem{schuld2018quantum}
Maria Schuld and Francesco Petruccione.
\newblock Quantum ensembles of quantum classifiers.
\newblock {\em Scientific Reports}, 8(1):2772, 2018.

\bibitem{kernelsmaria}
Maria Schuld and Nathan Killoran.
\newblock Quantum machine learning in feature hilbert spaces.
\newblock {\em Physical Review Letters}, 122(4):040504, 2019.

\bibitem{kubler2019quantum}
Jonas~M K{\"u}bler, Krikamol Muandet, and Bernhard Sch{\"o}lkopf.
\newblock Quantum mean embedding of probability distributions.
\newblock {\em arXiv preprint arXiv:1905.13526}, 2019.

\bibitem{havlivcek2019supervised}
Vojt{\v{e}}ch Havl{\'\i}{\v{c}}ek, Antonio~D C{\'o}rcoles, Kristan Temme,
  Aram~W Harrow, Abhinav Kandala, Jerry~M Chow, and Jay~M Gambetta.
\newblock Supervised learning with quantum-enhanced feature spaces.
\newblock {\em Nature}, 567(7747):209, 2019.

\bibitem{benedetti2018quantum}
Marcello Benedetti, John Realpe-G{\'o}mez, and Alejandro Perdomo-Ortiz.
\newblock Quantum-assisted helmholtz machines: a quantum--classical deep
  learning framework for industrial datasets in near-term devices.
\newblock {\em Quantum Science and Technology}, 3(3):034007, 2018.

\bibitem{4variational}
Alberto Peruzzo, Jarrod McClean, Peter Shadbolt, Man-Hong Yung, Xiao-Qi Zhou,
  Peter~J Love, Al{\'a}n Aspuru-Guzik, and Jeremy~L O’brien.
\newblock A variational eigenvalue solver on a photonic quantum processor.
\newblock {\em Nature Communications}, 5:4213, 2014.

\bibitem{romero2017quantum}
Jonathan Romero, Jonathan~P Olson, and Alan Aspuru-Guzik.
\newblock Quantum autoencoders for efficient compression of quantum data.
\newblock {\em Quantum Science and Technology}, 2(4):045001, 2017.

\bibitem{farhi2014quantum}
Edward Farhi, Jeffrey Goldstone, and Sam Gutmann.
\newblock A quantum approximate optimization algorithm.
\newblock {\em arXiv preprint arXiv:1411.4028}, 2014.

\bibitem{verdon2017quantum}
Guillaume Verdon, Michael Broughton, and Jacob Biamonte.
\newblock A quantum algorithm to train neural networks using low-depth
  circuits.
\newblock {\em arXiv preprint arXiv:1712.05304}, 2017.

\bibitem{schuld2018supervised}
Maria Schuld and Francesco Petruccione.
\newblock {\em Supervised Learning with Quantum Computers}, volume~17.
\newblock Springer, 2018.

\bibitem{DeutschJozsa}
D.~{Deutsch} and R.~{Jozsa}.
\newblock {Rapid Solution of Problems by Quantum Computation}.
\newblock {\em Proceedings of the Royal Society of London Series A},
  439:553--558, December 1992.

\bibitem{simon1997power}
Daniel~R Simon.
\newblock On the power of quantum computation.
\newblock {\em SIAM Journal on Computing}, 26(5):1474--1483, 1997.

\bibitem{schapire2003boosting}
Robert~E Schapire.
\newblock The boosting approach to machine learning: An overview.
\newblock In {\em Nonlinear estimation and classification}, pages 149--171.
  Springer, 2003.

\bibitem{casella2004generalized}
George Casella, Christian~P Robert, Martin~T Wells, et~al.
\newblock Generalized accept-reject sampling schemes.
\newblock In {\em A Festschrift for Herman Rubin}, pages 342--347. Institute of
  Mathematical Statistics, 2004.

\bibitem{schuldbook}
Maria Schuld and Francesco Petruccione.
\newblock Supervised learning with quantum computers, 2018.

\bibitem{dynamicprogramming}
Richard Bellman.
\newblock Dynamic programming.
\newblock {\em Science}, 153(3731):34--37, 1966.

\bibitem{highdimconvexity}
Bo’az Klartag.
\newblock High-dimensional distributions with convexity properties.
\newblock {\em Preprint, http://www. math. tau. ac. il/klartagb/papers/euro.
  pdf}, 2010.

\bibitem{beyer1999nearest}
Kevin Beyer, Jonathan Goldstein, Raghu Ramakrishnan, and Uri Shaft.
\newblock When is “nearest neighbor” meaningful?
\newblock In {\em International conference on database theory}, pages 217--235.
  Springer, 1999.

\bibitem{zimek2012survey}
Arthur Zimek, Erich Schubert, and Hans-Peter Kriegel.
\newblock A survey on unsupervised outlier detection in high-dimensional
  numerical data.
\newblock {\em Statistical Analysis and Data Mining: The ASA Data Science
  Journal}, 5(5):363--387, 2012.

\bibitem{gulde2003implementation}
Stephan Gulde, Mark Riebe, Gavin~PT Lancaster, Christoph Becher, J{\"u}rgen
  Eschner, Hartmut H{\"a}ffner, Ferdinand Schmidt-Kaler, Isaac~L Chuang, and
  Rainer Blatt.
\newblock Implementation of the deutsch--jozsa algorithm on an ion-trap quantum
  computer.
\newblock {\em Nature}, 421(6918):48, 2003.

\bibitem{nielsen2002quantum}
Michael~A Nielsen and Isaac Chuang.
\newblock Quantum computation and quantum information, 2002.

\bibitem{zhang}
Cha Zhang and Yunqian Ma.
\newblock {\em Ensemble machine learning: methods and applications}.
\newblock Springer, 2012.

\bibitem{berry}
Karl~O Friedrich et~al.
\newblock A berry-esseen bound for functions of independent random variables.
\newblock {\em The Annals of Statistics}, 17(1):170--183, 1989.

\bibitem{zwet}
Willem~R van Zwet.
\newblock A berry-esseen bound for symmetric statistics.
\newblock {\em Zeitschrift f{\"u}r Wahrscheinlichkeitstheorie und verwandte
  Gebiete}, 66(3):425--440, 1984.

\bibitem{convergence}
R.~N. Bhattacharya.
\newblock Refinements of the multidimensional central limit theorem and
  applications.
\newblock {\em The Annals of Probability}, 5(1):1--27, 1977.

\bibitem{shaft2006theory}
Uri Shaft and Raghu Ramakrishnan.
\newblock Theory of nearest neighbors indexability.
\newblock {\em ACM Transactions on Database Systems (TODS)}, 31(3):814--838,
  2006.

\end{thebibliography}

\section{Appendix A:}

\subsection{Analysis of the accuracy-weighted q-ensemble}
\noindent
Using the dequantised accuracy-weighted q-ensemble, we may study the behaviour of the algorithm. A problem occurs if it becomes difficult to select models with an accuracy greater than $0.5$ for the ensemble. In particular, if models tend to an accuracy of $0.5$ and the variance of these accuracies vanishes exponentially fast, both classical and quantum versions of the accuracy-weighted q-ensemble become intractable: exponentially more precision will be required to calculate the accuracies for the classical rejection criteria and the probabilities of the label in the quantum setting. Postselection is done on the $\ket{0}$-branch of the accuracy qubit in superposition $\ket{A_\theta} = \sqrt{a_\theta}\ket{0}+\sqrt{1-a_\theta}\ket{1}$. Looking at the probabilities, $p(\tilde{y}=-1) = \sum_{\theta \in \mathcal{E}^+} \frac{a_\theta}{E_{\mathcal{X}}}$ and
$p(\tilde{y}=1) = \sum_{\theta \in \mathcal{E}^-} \frac{a_\theta}{E_{\mathcal{X}}}$, if $a_\theta \to 0.5$ $\forall$ $\theta$ as $d \to \infty$ with increasing certainty, one requires increasingly more resources to obtain the necessary precision to compute the probabilities of the label. The resources needed are directly dependent on the rate of convergence of models' accuracies to $0.5$ and ultimately determine the feasibility of this strategy. In this analysis, we examine this rate of convergence with linear, perceptron and neural network models under specific data assumptions. \\

\subsubsection{Data and model assumptions}
\noindent We explicitly define the accuracy of a model with a parameter set $\theta$ as
\begin{equation}\label{acceqn}
    a_\theta = \frac{1}{M} \sum_{m=1}^{M} \frac{1}{2}|f(x_m;\theta) + f^{*}(x_m;\theta^{*})|,
\end{equation}

\noindent where $f$ is a base model, $f^*$ is a specified ``ground truth'' model with parameters $\theta^*$ and $a_\theta$ may be interpreted as the proportion of correct classifications of $M$ data points in a dataset $\mathcal{D}$ containing $d$ features. We can also define the average ensemble accuracy of $n$ sampled models for a fixed dimension (i.e. a fixed number of features) $d$ as $A_{\theta|d} = \frac{1}{n} \sum_{i=1}^{n} a_{\theta_i}$. To examine the behaviour of model accuracy we consider base models with simple data to provide an indication as to whether the accuracy-weighted strategy is indeed viable. The entries of the input vectors $x_m$ are drawn from a standard normal distribution and $\ell_2$ normalised. Again, for simplicity, the labels are assigned based on the first entry of each input vector. If the first entry of the $m^{\mathrm{th}}$ input vector denoted by $x_m^{(1)}$ is positive, $f^{*}(x_m;\theta^{*}) = 1$, else $f^{*}(x_m;\theta^{*}) = -1$. 

Model parameters $\theta$ contain entries drawn from a random uniform distribution over the interval $[ \ -1,1] \ $ to create the respective ensembles. This follows directly from the accuracy-weighted q-ensemble routine in \cite{schuld2018quantum} which assumes $\theta \in \ [a,b\ ]$ and initially encodes the parameter sets into a uniform superposition of computational basis states. Simulations are done with $M = 10 000$ input vectors and labels, along with $n = 100$ parameter sets. 

\begin{figure}
\centering
    \includegraphics[width=0.45\textwidth]{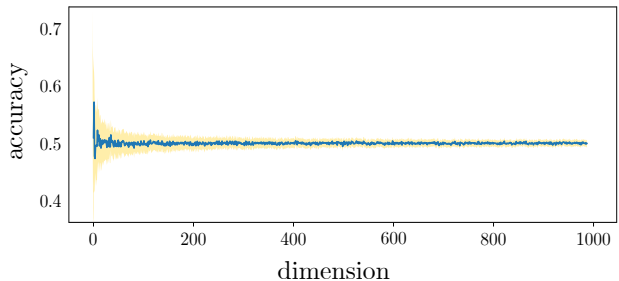}
    \caption{This plot contains the average accuracy of sampled parameter sets for linear models over increasing dimensions $d$. As $d$ increases, it becomes more noticeable that models in the ensemble have accuracies close to 0.5 with declining dispersion represented by one standard deviation above and below the mean in yellow shading. \label{linear}}
\end{figure}

\subsubsection{Accuracy in higher dimensions}

\begin{figure}
  \centering
  \includegraphics[width=.45\textwidth]{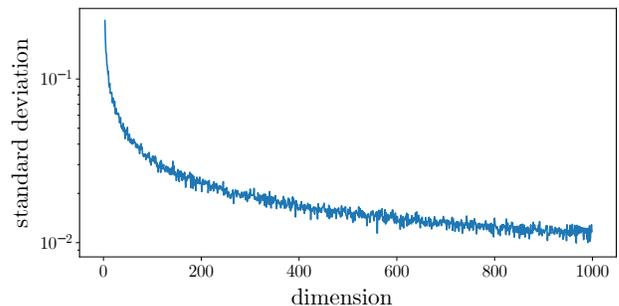}
  \caption{Declining standard deviation of average accuracies for linear models plotted on a log scale per dimension.}
  \label{LMvol}
\end{figure}

\begin{figure}
\centering
  \centering
  \includegraphics[width=.45\textwidth]{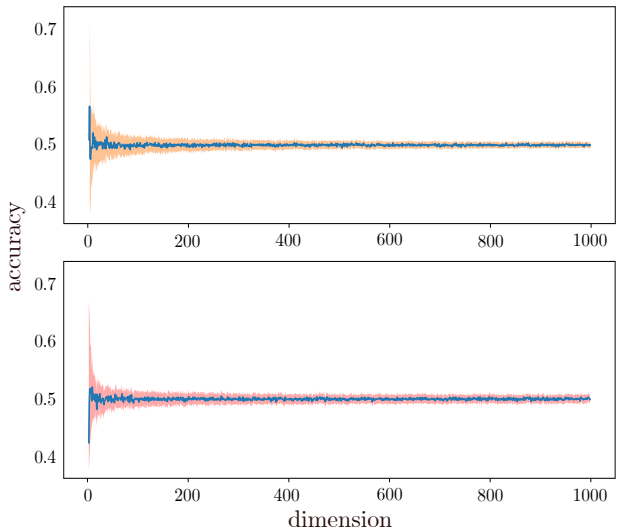}
  \caption{The average accuracy of sampled parameter sets for ensembles over increasing dimensions $d$ is depicted. Ensembles containing perceptron models are displayed in the top graph and neural networks with three hidden layers are contained in the bottom graph. The dispersion of model accuracies is represented by one standard deviation above and below the mean accuracies in the shaded orange and pink areas for perceptrons and neural networks respectively.}
  \label{percacc}
\end{figure}

\begin{figure}
  \centering
  \includegraphics[width=.45\textwidth]{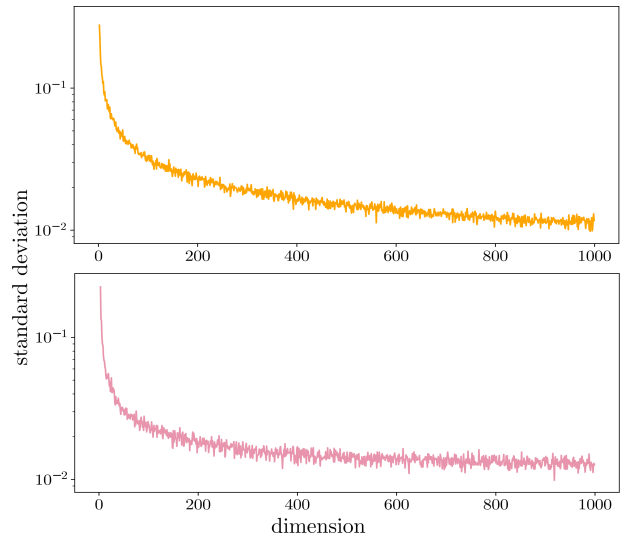}
  \caption{Declining standard deviation of average accuracies per dimension $d$ is plotted on a log scale for perceptron models in the top graph (in orange) and neural networks with three hidden layers in the bottom graph (in pink).}
  \label{NNvol}
\end{figure}

\noindent Figure \ref{linear} plots $A_{\theta|d}$ for $d \leq 1000$ using linear models and Figure \ref{percacc} contains the same plot for single layer perceptron models and neural networks with three hidden layers. These models all satisfy the point symmetric property assumed in the original accuracy-weighted q-ensemble and are commonly deployed in machine learning. For simulations in low dimensions ($d < 10$) the distribution of $A_{\theta|d}$ has a relatively high dispersion around the mean value of $0.5$ with all base models. This is desirable as it allows for models with accuracy $a_\theta > 0.5$ to be easily accepted into the ensemble. As the dimension increases, however, the dispersion rapidly decreases and all models tend to an accuracy of 0.5 with less deviation, making it increasingly difficult to build a strong ensemble. The sharp decline in standard deviation is plotted on a log scale for linear models in Figure \ref{LMvol} and Figure \ref{NNvol} for the perceptron and neural network models. This occurrence is due to the sampling method for the parameter vectors and the $\ell_2$ normalisation of data which leads to ``the curse of dimensionality'' and ``concentration of measure'' phenomena discussed in \cite{dynamicprogramming,highdimconvexity, beyer1999nearest}. Put briefly, most models become random guessers as the dimension increases.

Since $A_{\theta|d}$ is an average of the sum of random variables $a_{\theta}$, by the central limit theorem, $A_{\theta|d}$ tends to a normal distribution as more random variables are added. The mean and standard deviation of this distribution depend on $a_\theta$ which depends on the dimension of the data. We show in Appendix B that the mean of $a_\theta$ is equal to $0.5$ and the standard deviation tends to $0$ as $d$ increases to infinity for perceptron and linear models. Numerical simulations suggest the same behaviour for three layer neural networks as seen in Figure \ref{percacc}. The crucial issue is the rate of convergence of $a_\theta$ (and hence $A_{\theta|d}$) to $0.5$ which depends on the statistical properties of the input data. For $i.i.d$ data drawn from a standard normal distribution as done so here, by the Berry–Esseen theorem, the rate of convergence is $\mathcal{O}(d^{-\frac{1}{2}})$ \cite{berry,zwet}. For more realistic data, the convergence rate may differ. When data is represented as matrices, an expository survey of rates of convergence and asymptotic expansions in the context of the multi-dimensional central limit theorem may be found in \cite{convergence}. In short, the rates of convergence vary and are generally calculated (or estimated) on a case by case basis unless specific assumptions around the input data may be made. For the simple models deployed in this analysis, overall the results indicate convergence rates that are polynomial in $d$ and hence, the accuracy-weighted q-ensemble remains tractable. This, however, does not say anything about whether the accuracy-weighted q-ensemble (even if computationally tractable) is a useful method. As such, we continue with the same data assumptions, making use of the known tractability, and build the accuracy-weighted ensemble for a high-dimensional dataset.

\subsubsection{Poor performance of the accuracy-weighted ensemble on simple high-dimensional data}

\begin{figure}
    \centering
    \includegraphics[width=.45\textwidth]{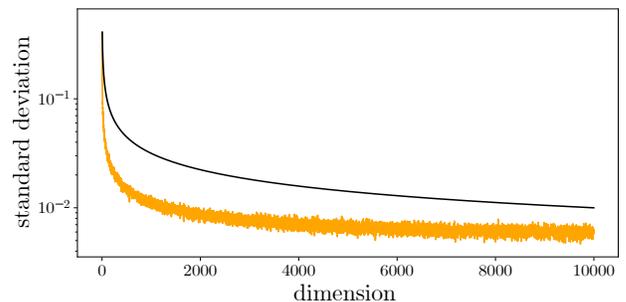}
    \caption{This plot contains the standard deviation of average accuracies of perceptron models in orange plotted on a log scale per dimension. The black line is the Berry-Essen asymptotic convergence.}
    \label{fig:mkclass}
\end{figure}

\noindent Using the perceptron as a base model, numerical simulations of the standard deviation of accuracies indicate that the rate of convergence is faster than the Berry-Esseen asymptotic convergence for $d \leqslant 10000$ as plotted in Figure \ref{fig:mkclass}. The standard deviation appears to level off for $d \geq 8000$ and the rate of decay is notably not exponential. Thus, one may build an ensemble of weak classifiers in higher dimensions. 

We construct the ensemble for $d = 8000$ to determine its empirical success using $M = 10 000$ training data points. With $n = 30 000$ sampled models, $14903$ models achieved an accuracy greater than $0.5$. These models were then selected for the ensemble and tested on $2000$ test samples. The overall test accuracy was found to be $0.50167$. Using $n = 100 000$ sampled models did not appear to improve the result as $49 580$ models passed the selection criteria and provided a test accuracy of $0.50165$. It is interesting that even with non-exponential convergence of the standard deviation to zero, the accuracy-weighted ensemble strategy reveals poor performance on a simple dataset that should be easy to classify. On the other hand, this does not imply that the accuracy-weighted method will perform poorly for all high-dimensional datasets. If, for example, data actually follows a mixture of distributions where classes are composed of natural clusters that each follow their own distribution, the curse of dimensionality may no longer be an issue \cite{beyer1999nearest,zimek2012survey,shaft2006theory}. 

\section{Appendix B:}
 \subsection{Modelling the behaviour of the perceptron ensemble in higher dimensions} 
 
 \begin{table}[H]
\centering
\begin{tabular}{c|c|r}
$\sigma(X_{ij}\theta_j)$ & $\sigma(X_{ij}\theta_j^*)$ & $\frac{1}{2}|\sigma(X_{ij}\theta_j) + \sigma(X_{ij}\theta_j^*)|$ \\ [0.5ex] \hline 
$+ 1$ & $+ 1$ & $1$ \\
$+ 1$ & $- 1$ & $0$ \\
$- 1$ & $+ 1$ & $0$ \\
$- 1$ & $- 1$ & $1$ \\
\end{tabular}
\caption{\label{tab:truth}Truth table for the expected outcomes of the perceptron model.}
\end{table}
 
 \noindent The perceptron model may be written as follows
 \begin{equation}\label{percep}
 f(\mathbf{X,\theta}) = \sigma(\mathbf{X\theta}),
 \end{equation}
 
 \noindent where $\sigma$ is a thresholding function ensuring the prediction is $1$ if the output is positive and $-1$ otherwise. $\mathbf{X}$ is a $M$ x $d$ matrix where each row represents one data point and $\theta$ is a $d$ x $1$ column vector. The ground truth model in this example may be formulated with $\theta^* = \ [ 1, 0, ... , 0\ ]^T$ for simplicity, although this may be any uniformly sampled $\theta$. One may now rewrite the accuracy of a model from Equation (\ref{acceqn}) in terms of expectation values
 
 \[
 \mathrm{E}\ [a_{\theta} \ ] = \frac{1}{M} \sum_{i=1}^M \sum_{j=1}^d \frac{1}{2} |\sigma(X_{ij}\theta_j) + \sigma(X_{ij}\theta_j^{*})|.
 \]

\noindent Since $X_{ij} \sim N(0,1)$ and $\theta_j \sim U\ [-1,1\ ]$, $\sigma(X_{ij}\theta_j) = \pm 1$ with equal probability, and similarly for $\sigma(X_{ij}\theta^*)$. Looking at all possible outcomes presented in Table \ref{tab:truth}, the distribution of accuracy is $1$ and $0$ with approximately equal occurrence. As $d \to \infty$, by the law of large numbers, the distribution of accuracy contains an equal number of $0$'s and $1$'s with more certainty. Taking the expectation of this gives exactly $0.5$ as confirmed by the simulated results
\begin{align}
    \lim_{d\to\infty} \mathrm{E}\ [a_{\theta} \ ] &= \frac{1}{M} \sum_{i=1}^M \sum_{j=1}^d \frac{1}{2} |\sigma(X_{ij}\theta_j) + \sigma(X_{ij}\theta_j^{*})| \\
    &= \frac{1}{M} \sum_{i=1}^M \frac{1}{2} = \frac{1}{2}. \label{accresulteqn}
\end{align}

\noindent Using a similar logic, the variance of the accuracy of the perceptron model may be formulated as follows

\begin{align}\label{varianceeqn}
    \mathrm{Var}\ [a_\theta\ ] &= \mathrm{E}\ [a_\theta^2 \ ] - \mathrm{E}\ [a_\theta \ ]^2.
\end{align}

\noindent The thresholding function applies to all $f(\mathbf{X},\theta)$ and so we leave it out of the notation for brevity but its effect is still present. The expectation of the squared accuracy is 

\begin{equation*}
\setlength{\jot}{10pt}
\begin{split}
    \mathrm{E}[a_\theta^2 \ ] = \frac{1}{M} \sum_{i=1}^M \sum_{j=1}^d \left (\frac{1}{2} |X_{ij}\theta_j + X_{ij}\theta_j^{*}|\right )^2, \\
    = \frac{1}{M} \sum_{i=1}^M \sum_{j=1}^d  (\frac{1}{4}X_{ij}\theta_jX_{ij}\theta_j + \frac{1}{4}X_{ij}\theta_j^*X_{ij}\theta_j^* \\ +  \frac{2}{4}X_{ij}\theta_jX_{ij}\theta_j^*).
\end{split}
\end{equation*}

\noindent Taking the expectation of each of the terms, it is clear that the first and last terms equal $0$ as $d$ increases to infinity since $X_{ij}$ and $\theta_j$ are independent with mean values of $0$. The middle term may be further simplified using the fact that $\theta^* =\ [1, 0, ..., 0\ ]^T$

\begin{align*}
\setlength{\jot}{10pt}
    \lim_{d \to \infty} \mathrm{E}\ [a_\theta^2 \ ] &= \frac{1}{M} \sum_{i=1}^M \sum_{j=1}^d \frac{1}{4}X_{ij}\theta_j^*X_{ij}\theta_j^*, \\
    &= \frac{1}{M} \sum_{i=1}^M \frac{1}{4}X_{i1}^2, \\
    &= \frac{1}{M} \sum_{i=1}^M \frac{1}{4}(1), \\
    &= \frac{1}{4}.
\end{align*}

\noindent Using this along with the result from Equation (\ref{accresulteqn}), the variance from Equation (\ref{varianceeqn}) becomes

\begin{align*}
    \lim_{d \to \infty}\mathrm{Var}\ [a_\theta\ ] &= \mathrm{E}\ [a_\theta^2 \ ] - \mathrm{E}\ [a_\theta \ ]^2, \\
    &= \frac{1}{4} - \left(\frac{1}{2}\right)^2, \\
    &= 0,
\end{align*}
\noindent which holds for any uniformly sampled ground truth model.

\end{document}